\documentclass[aps,letterpaper,10pt,twocolumn,floats,showpacs,amsmath,amsfonts,amssymb,pre]{revtex4-1}

\usepackage{graphicx}% Include figure files
\usepackage[colorlinks=true,citecolor=blue]{hyperref}
\usepackage[caption=false]{subfig}
\usepackage{pstricks}
\usepackage{amsmath}
\begin{document}

\preprint{APS/123-QED}

\title{First-passage times in renewal and nonrenewal systems}% Force line breaks with \\

\author{Krzysztof Ptaszy\'{n}ski}
 \email{krzysztof.ptaszynski@ifmpan.poznan.pl}
\affiliation{%
 Institute of Molecular Physics, Polish Academy of Sciences, ul. M. Smoluchowskiego 17, 60-179 Pozna\'{n}, Poland
}%

\date{\today}% It is always \today, today,

\begin{abstract}
Fluctuations in stochastic systems are usually characterized by the full counting statistics, which analyzes the distribution of the number of events taking place in the fixed time interval. In an alternative approach, the distribution of the first-passage times, i.e. the time delays after which the counting variable reaches a certain threshold value, is studied. This paper presents the approach to calculate the first-passage time distribution in systems in which the analyzed current is associated with an arbitrary set of transitions within the Markovian network. Using this approach it is shown that when the subsequent first-passage times are uncorrelated there exist strict relations between the cumulants of the full counting statistics and the first-passage time distribution. On the other hand, when the correlations of the first-passage times are present, their distribution may provide additional information about the internal dynamics of the system in comparison to the full counting statistics; for example, it may reveal the switching between different dynamical states of the system. Additionally, I show that breaking of the fluctuation theorem for first-passage times may reveal the multicyclic nature of the Markovian network.

\end{abstract}

%05.40.-a Fluctuation phenomena, random processes, noise, and Brownian motion
%72.70.+m	Noise processes and phenomena
%73.23.Hk	Coulomb blockade; single-electron tunneling

\maketitle

%\tableofcontents

\section{\label{sec:intro}Introduction}
Statistics of fluctuations in stochastic systems provide important information about their dynamics and thermodynamics. The noteworthy example of this fact are the fluctuation theorems, like these of Jarzynski~\cite{jarzynski1997} or Crooks~\cite{crooks1999}, which describe the universal properties of nonequilibrium fluctuations of thermodynamic quantities (cf. the review paper by Seifert~\cite{seifert2012}). Apart from their fundamental importance in nonequilibrium thermodynamics, they have been experimentally applied to reconstruct the free energy landscape of molecules~\cite{libhardt2002, harris2007, gupta2011}. The other example are the universal thermodynamic bounds on cumulants of current fluctuations~\cite{barato2015prl, barato2015, gingrich2016, pietzonka2016, polettini2016, pietzonka2017, horowitz2017}. Fluctuations may also provide information about the details of the internal dynamics. For example, cumulants of current fluctuations provide a bound on the minimal number of states within the Markovian network~\cite{schnitzer1995, barato2015prlb}. In some cases, they may even enable the whole reconstruction of the generator of the dynamics of the system~\cite{bruderer2014}. Analysis of current fluctuations has been already applied to the study of transport mechanism in electronic systems~\cite{bulka2000, belzig2005, koch2005, gustavsson2006, gustavsson2006b, fricke2007, sothmann2010, ubbelohde2013, cascales2014, lau2016}, biomolecular kinetics~\cite{schnitzer1995, schnitzer1997, shaevitz2005, choi2012} or dynamics of photon emitters~\cite{matthiesen2014}.

Statistics of fluctuations can be divided into the fixed time and the fluctuating time statistics~\cite{budini2014, esposito2010}. To the former category belongs the full counting statistics~\cite{bagrets2003, touchette2009, wachtel2015}, which determines the probability distribution of the number of event taking place in a given time interval. Usually the zero-frequency counting statistics is analyzed, which assumes the time interval tending to infinity. To the latter category belongs, for example, the waiting time distribution~\cite{brandes2008}, which determines the probability density of the time delays between events of the same type.   

One may ask, if the fixed time and the fluctuating time statistics are equivalent, or if they provide a distinct information. If the observed process can be described by the renewal theory, which assumes that the waiting times between the subsequent events are uncorrelated, the former appears to be true. For example, there exist strict relations between cumulants of the zero-frequency full counting statistics and the waiting time distribution~\cite{albert2011, budini2011}. In Markovian systems all elementary processes, i.e. transitions between single states of the network, are always renewal. However, the observed events (like the electron jumps, photon emissions, steps of the molecular motor etc.) are often associated with a set of two or more elementary transitions; in such a case, the analyzed process may exhibit a nonrenewal behavior. Such a situation has been already investigated for the cases of enzymatic reactions~\cite{gopich2006, saha2011}, molecular motors~\cite{bhat2013}, emission of photons by fluctuating emitters~\cite{cao2006, caycedo-soler2008, budini2010, osadko2011} or Josephson junctions~\cite{dambach2015}, electron transport through quantum dots~\cite{ptaszynski2017, ptaszynski2017b} and molecules~\cite{kosov2017} or statistics of neuronal spike trains~\cite{akerberg2011}. When the renewal theory does not longer hold, no obvious relation between the fixed time and the fluctuating time statistics exists and thus both approaches may provide a complementary information~\cite{ptaszynski2017}. It has been recently shown, by using the example of the double quantum dot molecule, that in nonrenewal systems the waiting time distribution may give information about the internal dynamics which cannot be provided by the zero-frequency full counting statistics at all~\cite{ptaszynski2017b}. This highlights the usefulness of the analysis of the fluctuating time statistics.

However, while the waiting time distribution is well suited for the analysis of currents associated with unidirectional transitions between states of the Markovian network, it does not give consistent description of currents associated with bidirectional transitions. Bidirectionality of transitions is, on the other hand, required for the thermodynamical consistency within the formalism of stochastic thermodynamics~\cite{seifert2012}. As a matter of fact, many physical processes observed in nature, like motion of molecular motors~\cite{nishiyma2002} or electron tunneling through quantum dots~\cite{fujisawa2006}, involve bidirectional transitions. Fluctuating time statistics of bidirectional transitions can be, however, characterized by using the first-passage time distribution~\cite{bauer2014, roldan2015, saito2016, neri2017, garrahan2017}. In this approach the distribution of the time delays after which the observed quantity (for example the number of transitions) reaches some threshold value is analyzed. Since the observed quantity can be defined as a difference of the number of transitions in different directions, the bidirectional processes can be consistently investigated. This enables, for example, to derive fluctuation theorems relating the first-passage time distribution to the entropy production~\cite{bauer2014, roldan2015, saito2016, neri2017}.

It is then natural to ask, if the first-passage time distribution is in some way related to the fixed time statistics. This paper gives an answer to this question. I investigate the first-passage times in a discrete Markovian network using the approach introduced by Saito and Dhar~\cite{saito2016}, generalized by me in Sec.~\ref{sec:method} to describe both the renewal and the nonrenewal systems. In Sec.~\ref{sec:singleres} for a special class of single-reset system, which are known to exhibit a renewal behavior, the relations between cumulants of the first-passage time distribution and the full counting statistics are derived, which generalize the previously known relations between the full counting statistics and the waiting time distribution. I also provide a heuristic argumentation for the generality of these relations for an arbitrary renewal system, which is relegated to Appendix~\ref{sec:appendixb}. Section~\ref{sec:nonrenewal} shows that the derived formulas do not longer hold in the case of nonrenewal systems, and present a way to characterize the correlation between subsequent first-passage times. In~Sec.~\ref{sec:fluctuation} the difference between the unicyclic and multicyclic systems is briefly discussed by analyzing the validity of the fluctuation theorem for the first-passage times. Section~\ref{sec:conclusions} brings conclusions following from my results. Appendix~\ref{sec:appendix} includes some mathematical details.

\section{\label{sec:method}Calculation of the first-passage time distribution}
I consider a Markovian network consisting of $M$ discrete states ${i}$ connected to $K$ thermal reservoirs ${\alpha}$ with corresponding temperatures $T_\alpha$. The transition rate from the state $j$ to the state $i$ is denoted as $k_{ij}$. Each transition rate can be further decomposed into a sum of rates corresponding to different reservoirs: $k_{ij}=\sum_{\alpha} k_{ij}^{\alpha}$. When each transition associated with each reservoir is reversible (for each $k_{ij}^{\alpha}$ the condition $k_{ij}^{\alpha}/k_{ji}^{\alpha}>0$ is fulfilled) the model is thermodynamically consistent~\cite{seifert2012}. The dynamics of the system can be described by the master equation
\begin{eqnarray} \label{mastereq}
\dot{\mathbf{p}}(t)=\mathbf{W} \mathbf{p}(t),
\end{eqnarray}
where $\mathbf{p}(t)$ is the vector of state probabilities $p_i(t)$ and $\mathbf{W}$ is the matrix containing the elements $W_{ij}=k_{ij}$ for $i \neq j$ and $W_{ii}=-\sum_{j\neq i} k_{ji}$. Here I focus on systems in which the quantum coherence is absent; however the method can be easily generalized to describe the coherent systems described by means of the quantum master equation~\cite{carmichael1993, breuer2002}.

Let us now define two operators $\mathbf{J}_F$ and $\mathbf{J}_B$ corresponding to two different sets of transitions $k_{ij}^\alpha$, later referred to as ``forward'' and ``backward'' transitions. These operators may correspond, for example, to electrons jumps to/from a chosen lead or to the steps of the molecular motor in the forward/backward directions. Let us also define the jump number $n=n_F-n_B$, where $n_F$ ($n_B$) is a number of forward (backward) transitions occurring in the time interval $[0,t]$. The first-passage time distribution $F(N|\tau)$ is then defined as the probability density, that the jump number $n$ reaches the value $N$ in the moment $\tau$ for the first time. This function depends on the initial state described by the vector $\mathbf{p}(0)$.

To determine $F(N|\tau)$ the following procedure, developed by Saito and Dhar~\cite{saito2016}, is used. First, the vector $\mathbf{p}(t)$ is decomposed into a sum of vectors corresponding to a specific value of the jump number:
\begin{align}
\mathbf{p}(t) = \sum_{n=-\infty}^\infty \mathbf{p}^{(n)}(t).
\end{align}
It is also useful to define the generating function
\begin{align}
\mathbf{p}(z,t) = \sum_{n=-\infty}^\infty z^n \mathbf{p}^{(n)}(t),
\end{align}
where $z$ is a complex number. It is given by the following equation:
\begin{align}
\mathbf{p}(z,t) = e^{\mathbf{W}_z t} \mathbf{p}(0),
\end{align}
which is a solution of the equation $\dot{\mathbf{p}}(z,t) = \mathbf{W}_z \mathbf{p}(z,t)$ with $\mathbf{W}_z=\mathbf{W}-\mathbf{J}_F-\mathbf{J}_B+\mathbf{J}_F z +\mathbf{J}_B/ z$. The $n$-conditioned probability vector $\mathbf{p}^{(n)}(t)$ can be then written as
\begin{align} \label{condvec}
\mathbf{p}^{(n)}(t)=\frac{1}{2 \pi i} \oint \frac{dz}{z^{n+1}} \mathbf{p}(z,t)=\mathbf{T}(n|t) \mathbf{p}(0),
\end{align}
where the integration goes along the unit circle around 0. $\mathbf{T}(n|t)$ is the transition matrix defined in the following way:
\begin{align} \label{tmat}
\mathbf{T}(n|t)=\frac{1}{2 \pi i} \oint \frac{dz}{z^{n+1}} e^{ \mathbf{W}_z t}.
\end{align}
It is also useful to consider its Laplace transform:
\begin{align}  \label{tmatlapl}
\hat{\mathbf{T}}(n|s)&=\frac{1}{2 \pi i} \oint \frac{dz}{z^{n+1}} \frac{1}{s-\mathbf{W}_z} \\ \nonumber
&=\frac{1}{2 \pi i} \oint \frac{dz}{z^{n+1}} \frac{\mathbf{C}(z,s)}{\det[s-\mathbf{W}_z]},
\end{align}
where $\mathbf{C}(z,s)$ is the cofactor matrix of the matrix ${s-\mathbf{W}_z}$.

Now, one can determine the first-passage time distribution. In the paper of Saito and Dhar~\cite{saito2016} the case, when the counted process is associated with the single transition between states of the Markovian network, i.e. the matrices $\mathbf{J}_F$ and $\mathbf{J}_B$ contain only one non-zero element, has been considered. Here the general case, when the jump operator is associated with an arbitrary set of transitions within the network, i.e. the matrices $\mathbf{J}_F$ and $\mathbf{J}_B$ may contain several non-zero elements in different rows and columns, is analyzed. As a matter of fact, in many physical systems the observed current is associated with such complex jump operators~\cite{gopich2006, brandes2008, gingrich2017, ptaszynski2017}. Let us consider how the jump-number-conditioned probability of the state $i$ in the moment $t$, denoted as $p_i^{(N)}(t)$, can be determined. First, according to Eq.~\eqref{condvec} one obtains
\begin{align} \label{prob1}
p_i^{(N)}(t)= \sum_{j} T_{ij}(N|t) p_j(0),
\end{align}
where $T_{ij}(N|t)$ is the element of the matrix $\mathbf{T}(N|t)$ (representing the transition from the $j$ to the $i$ state) whereas $p_j(0)$ is the element of the vector $\mathbf{p}(0)$. 

On the other hand, one may observe that in the time interval $[0,t]$ many different stochastic trajectories (i.e. sequences of the transitions between states of the Markovian network) can be realized. Let us use the following notation: a situation in which the state $k$ is occupied and the jump number equals $n$ is denoted as $(k,n)$. Probability of the state $(k,n)$ in the moment $t'$ is equal to $p_k^{(n)}(t')$. Each transition changes the state of the system, and may also change the jump number; such transitions are denoted as ${(k,n) \rightarrow (l,n')}$ where $n' \in \{n-1,n,n+1\}$. Without loss of generality, let us now consider the case of $N>0$. The state ${(i,N)}$ can be reached through different stochastic trajectories of the type
\begin{align} \label{trajectory}
(j,0) \rightarrow \dots \rightarrow (k,N-1) \rightarrow (l,N) \rightarrow \dots \rightarrow (i,N).
\end{align}
Here ${(k,N-1) \rightarrow (l,N)}$ denotes the transition in which the jump number reaches the threshold value $N$ \textit{for the first time}. Such trajectories can be then divided into different sets. Let us consider the set of trajectories for which some transition $(k,N-1) \rightarrow (l,N)$, with arbitrary $k$ but specific $l$, takes place in the moment $\tau$. The probability density that the trajectory belongs to such a set is equal to $F_l(N|\tau) T_{il}(0|t-\tau)$. The first factor, $F_l(N|\tau)$, is the probability density that the transition ${(k,N-1) \rightarrow (l,N)}$ (as above, with arbitrary $k$) takes place in the moment $\tau$; i.e. this is probability density that two conditions are met: the jump number reaches $N$ in the moment $\tau$ for the first time and this is associated with initialization of the state $l$. As the first-passage time distribution, the function $F_l(N|t)$ depends on the initial vector $\mathbf{p}(0)$. Summing such functions over all states $l$ one obtains the total first-passage time distribution:
\begin{align} \label{condfptd}
F(N|\tau)= \sum_l F_l (N|\tau).
\end{align}
The second factor, $T_{il}(0|t-\tau)$, is the conditional probability that if the state $l$ is occupied in the moment $\tau$ the state $i$ will be occupied in the moment $t$ without change of the jump number. The probability $p_i^{(N)}(t)$ is then the sum of terms $F_l(N|\tau) T_{il}(0|t-\tau)$ over all sets of trajectories, which can be expressed as
\begin{align} \label{prob2}
p_i^{(N)}(t)=\sum_{l } \int_0^t F_l (N|\tau) T_{il}(0|t-\tau) d\tau.
\end{align}
The same result can be derived for $N<0$ [with the transition ${(k,N-1) \rightarrow (l,N)}$ in Eq.~\eqref{trajectory} replaced by ${(k,N+1) \rightarrow (l,N)}$]. 

Comparing Eqs.~\eqref{prob1} and~\eqref{prob2}, and changing index $l \rightarrow j$ in Eq.~\eqref{prob2}, one obtains 
\begin{align} \label{uklad}
\sum_{j} T_{ij}(N|t) p_j(0)= \sum_{j } \int_0^t F_j (N|\tau) T_{ij}(0|t-\tau) d\tau.
\end{align}
One can write a system of such equations for different final states $i$. Solving such a system one may determine all functions $F_j(N|\tau)$, and then calculate the first-passage time distribution $F(N|\tau)$ using Eq.~\eqref{condfptd}. This is the first main result of the paper.

Equation~\eqref{uklad} is a Volterra equation of a convolution type, which can be solved using the Laplace transform
\begin{align} \label{ukladlapl}
& \sum_{j } \hat{T}_{ij}(N|s) p_j(0) = \sum_{j } \hat{F}_j (N|s) \hat{T}_{ij}(0|s).
\end{align}
Functions $\hat{F}_j (N|s)$ can be then obtained by solving a system of linear equations. Here and in the whole paper the ``hat'' symbol is used to denote the Laplace transform. Equation~\eqref{ukladlapl} can be written in the matrix form:
\begin{align}
 \hat{\mathbf{T}}(N|s) \mathbf{p}(0)=\hat{\mathbf{T}}(0|s) \hat{\mathbf{F}}(N|s),
\end{align}
where the column vector $\hat{\mathbf{F}}(N|s)$ is defined as $\hat{\mathbf{F}}={(\hat{F}_1, \hat{F}_2, \dots)^T}$. Multiplying both sides by $\hat{\mathbf{T}}(0|s)^{-1}$, tracing both sides and applying Eq.~\eqref{condfptd} one obtains the solution
\begin{align} \label{ukladlaplmat}
\hat{F}(N|s)= \text{Tr} \left[ \hat{\mathbf{T}}(0|s)^{-1} \hat{\mathbf{T}}(N|s) \mathbf{p}(0) \right],
\end{align}
which exhibits some similarity to the formula for the waiting time distribution derived by Brandes~\cite{brandes2008}. 

It is often useful to consider cumulants of the first-passage time distribution instead of the distribution itself. They can be calculated in the following way:
\begin{align} \label{kumfpt}
\kappa^N_m=(-1)^m  \lim_{s \rightarrow 0^{+}} \left[\frac{d^m}{ds^m} \ln \hat{F}(N|s) \right].
\end{align}
One should be aware that since the matrix $\mathbf{W}_z$ is singular for $z=1$ (because a sum of all elements in each column equals 0) the matrix $\hat{\mathbf{T}}(n|s)$, and therefore the right-hand side of Eq.~\eqref{ukladlaplmat}, is not defined for $s=0$ [cf.~Eq~\eqref{tmatlapl}]. This is why the right-sided limit is used in Eq.~\eqref{kumfpt}.

In practice, the most demanding part of the calculation is the evaluation of the integral over a complex variable $z$ in Eqs.~\eqref{tmat}-\eqref{tmatlapl}, which often requires the use of numerical methods. For the calculation of the cumulants it is convenient to expand the function ${\mathbf{C}(z,s)/\det[s-\mathbf{W}_z]}$ into the Taylor series about a very small but finite value of $s$, and then numerically integrate every element of the series over $z$ to obtain the Taylor expansion of $\hat{\mathbf{T}}(n|s)$. When the system is far from equilibrium, it is often sufficient to use only the first few elements of the series to achieve a good convergence of the low-order cumulants.

In general, the vector of the initial state $\mathbf{p}(0)$ can be chosen in an arbitrary way. However, to make the method comparable to the previously considered approaches, from this moment the following convention is used to define $\mathbf{p}(0)$: when one determines the distribution $F(N|\tau)$ for the positive threshold value ($N>0$) the measurement of the single first-passage time begins when some ``forward'' jump takes place (the initial jump is not yet counted) and stops when the jump number reaches $N$ due to another ``forward'' jump. This is analogous to the measurement of the waiting time distribution, in which one determines the time delays between the subsequent jumps~\cite{brandes2008}. In the same way, when $F(N|\tau)$ for $N<0$ is analyzed the measurement begins when some ``backward'' jump occurs and stops when the threshold $N$ is reached due to another ``backward'' transition. For such a convention, the vector of the initial state is defined as:
\begin{align} \label{definitial}
\mathbf{p}(0) =
\begin{cases}
\mathbf{J}_F \mathbf{p}_s/\text{Tr}(\mathbf{J}_F \mathbf{p}_s)  &  \text{for} \quad N>0, \\
\mathbf{J}_B \mathbf{p}_s/\text{Tr}(\mathbf{J}_B \mathbf{p}_s)  &  \text{for} \quad N<0,
\end{cases}
\end{align}	
where $\mathbf{p}_s$ is the vector of the stationary state (solution of the equation $\mathbf{W} \mathbf{p}_s=0$). When such a definition is used, in the case of unidirectional transitions ($\mathbf{J}_B=0$) the first-passage time distribution $F(1|\tau)$ is equivalent to the waiting time distribution defined by Brandes~\cite{brandes2008}.

\section{\label{sec:singleres}First-passage times in single-reset systems}
In this section I consider single-reset systems, defined as ones in which every ``forward'' jump leads to the initialization of the same state $\nu$. This means that matrix $\mathbf{J}_F$ contains non-zero elements only in the $\nu$th row and $F(N|\tau)=F_\nu(N|\tau)$ for $N>0$. As a matter of fact, many relevant systems, like quantum dots in the strong Coulomb blockade regime~\cite{brandes2008} or simple molecular motors and enzymatic networks~\cite{gopich2006, chemla2008}, are single-reset ones. In such systems, in the case of unidirectional transitions ($\mathbf{J}_B=0$), the waiting time distribution exhibits the renewal property -- the subsequent waiting times are uncorrelated~\cite{brandes2008, ptaszynski2017}. Here I prove that the same is true for the first-passage time distribution in the case of bidirectional transitions. Moreover, the relations between cumulants of the first-passage time distribution and the full counting statistics are derived; they are generalizations of the ones relating the cumulants of the full counting statistics and the waiting time distribution in the renewal systems which have been presented in Refs.~\cite{albert2011, budini2011}.

I focus on the situation when $N>0$ and the current in the forward direction is positive: $\text{Tr}[(\mathbf{J}_F-\mathbf{J}_B)\mathbf{p}_s]>0$. According to the applied convention [Eq.~\eqref{definitial}] state $\nu$ is the initial state (one should be aware that all results below are valid only when this convention is used). For the case analyzed, Eq.~\eqref{ukladlapl} takes the simple form
\begin{align} \label{integraleq1lapl}
\hat{F}(N|s)=\frac{\hat{T}_{\nu \nu}(N|s)}{\hat{T}_{\nu \nu}(0|s)},
\end{align}
where the element $T_{\nu \nu}(n|t)$ of the transition matrix reads
\begin{align} \label{tnnint}
\hat{T}_{\nu \nu}(n|s)& = \frac{1}{2 \pi i} \oint \frac{dz}{z^{n+1}} \left( \frac{1}{s-\mathbf{W}_z} \right)_{\nu \nu} \\ \nonumber &=\frac{1}{2 \pi i} \oint \frac{dz}{z^{n+1}} \frac{C_{\nu \nu}(z,s)}{\det[s-\mathbf{W}_z]}.
\end{align}

In the matrix $s-\mathbf{W}_z$ only the $\nu$th row and the $\nu$th column contain elements $z$ and $z^{-1}$, respectively. This has two important consequences. First, the cofactor $C_{\nu \nu}(z,s)=C_{\nu \nu}(s)$ is $z$-independent. Secondly, the determinant $\det[s-\mathbf{W}_z]$ contains only elements proportional to $z$, $z^0$ and $z^{-1}$, and therefore can be expressed in the following way:
\begin{align} \label{detpolyn}
\det[s-\mathbf{W}_z]=a(s)\frac{\left[ z-z_+(s)\right]\left[ z-z_-(s)\right]}{z},
\end{align}
where $a(s)$ is some function of $s$, whereas $z_+(s)$ and $z_-(s)$ are the higher and the smaller root of the equation $\det[s-\mathbf{W}_z]=0$. Derivation of Eq.~\eqref{detpolyn} is based on the properties of the determinants of the block matrices (see Appendix~\ref{sec:appendix}). Equation~\eqref{tnnint} can be then rewritten as
\begin{align} \label{tnnint2}
\hat{T}_{\nu \nu}(n|s)& = \frac{f(s)}{2 \pi i} \oint \frac{dz}{z^{n}\left[ z-z_+(s)\right]\left[ z-z_-(s)\right]},
\end{align}
where $f(s)=a(s) C_{\nu \nu}(s)$.

%%%%%%%%%%%%%%%%%%%%%%%%%%%%%%%%%%%%%%%%%%%%%%%%%%%%%%%%%%%%%%%%%%%%
\begin{figure}
	\centering
	\includegraphics[width=0.97\linewidth]{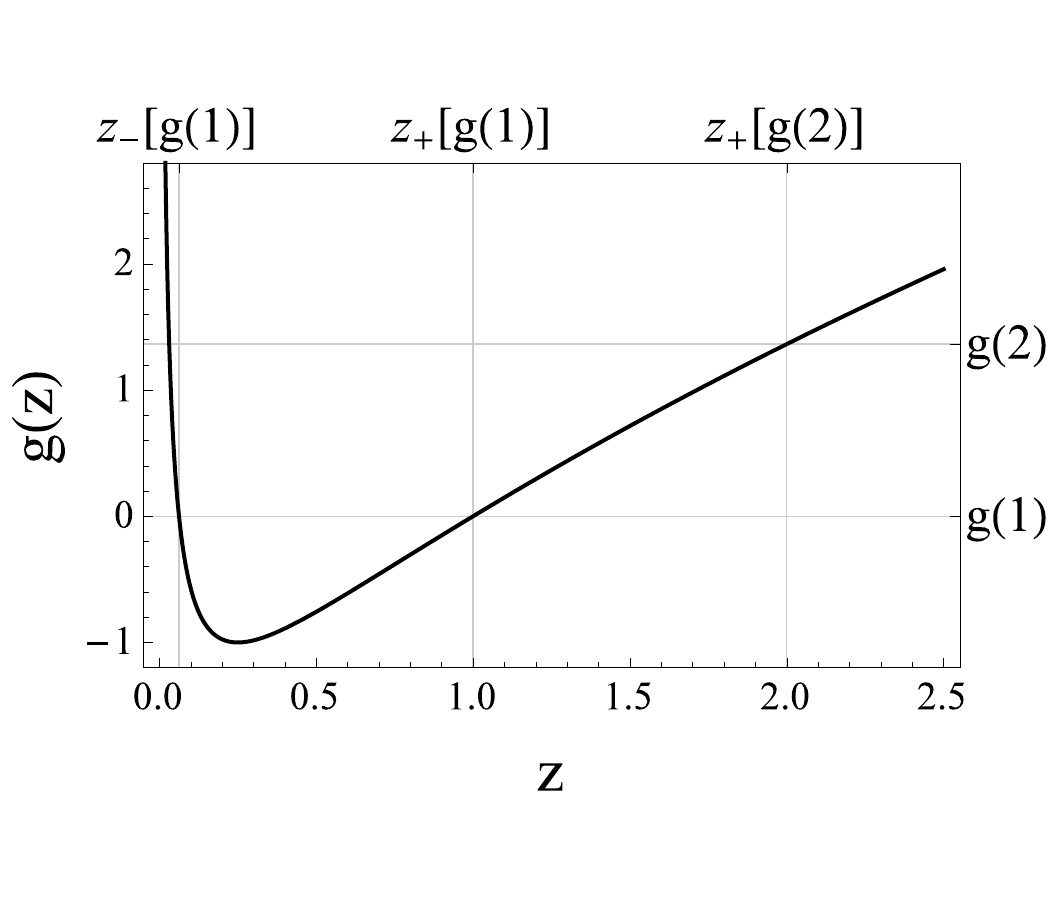}
	\caption{Plot of an exemplary function $g(z)$ illustrating the properties of the functions $z_{+}(s)$ and $z_{-}(s)$, i.e. the larger and the smaller root of the equation $g(z)=s$ [cf. discussion below Eq.~\eqref{scfungen}].}
	\label{fig:fungen}
\end{figure}
%%%%%%%%%%%%%%%%%%%%%%%%%%%%%%%%%%%%%%%%%%%%%%%%%%%%%%%%%%%%%%%%%%%%

On the other hand the equation
\begin{align} \label{scfungen}
\det[g(z)-\mathbf{W}_z]=0,
\end{align}
defines the scaled cumulant generating function $g(z)$, which, for systems having a unique stationary state, is a dominant eigenvalue of the matrix $\mathbf{W}_z$~\cite{touchette2009}. This function is more commonly expressed in the form $\lambda(l)=g(e^l)$; here its different form is used to simplify some derivations. Comparing Eqs.~\eqref{detpolyn} and~\eqref{scfungen} one finds that $z_{+}(s)$ and $z_{-}(s)$ are two roots of the equation $g(z)=s$: the equality $g(z)=s$ is satisfied only when the right-hand side of Eq.~\eqref{detpolyn} is equal to 0, i.e $z=z_{+}(s)$ or $z=z_{-}(s)$. This indicates that either the equality $z=z_{+}[g(z)]$ or $z=z_{-}[g(z)]$ must hold. Because for real $z$ the function $g(z)$ is convex and has a unique global minimum at $z<1$~\cite{touchette2009}, the smaller root of the equation $g(z)=s$ must be lower than one, i.e. $z_{-}(s)<1$. Therefore for $z \geq 1$ the relation $z=z_{+}[g(z)]$ holds. Furthermore, since $g(1)=0$~\cite{touchette2009} one finds $z_{+}[g(1)]=z_+(0)=1$. Due to convexity of the function $g(z)$ the relation $g(z)>0$ holds for $z >1$; in consequence $z_{+}(s)>1$ for $s>0$. These properties are illustrated in Fig.~\ref{fig:fungen}.

Using the properties $z_{+}(s)>1$ and $z_{-}(s)<1$ for $s>0$ one can solve the integral in Eq.~\eqref{tnnint2}. It is equal to the $(n-1)$th element of the Laurent series of the function $\left\{\left[ z-z_+(s)\right]\left[ z-z_-(s)\right]\right\}^{-1}$, which can be found directly by expanding $\left[ z-z_+(s)\right]^{-1}=-[1/z_+(s)] \sum_{n=0}^\infty [z/z_+(s)]^n$ and $\left[ z-z_-(s)\right]^{-1}=(1/z) \sum_{n=0}^\infty [z_-(s)/z]^n$. Equation~\eqref{tnnint2} can be then expressed in the simple form
\begin{align}
\hat{T}_{\nu \nu}(n|s)& = \frac{f(s) z_{+}(s)^{-n}}{z_{-}(s)-z_{+}(s)}.
\end{align}
Finally, using Eq.~\eqref{integraleq1lapl} one obtains
\begin{align} \label{fptdsingresodz}
\hat{F}(N|s)= [z_{+}(s)]^{-N},
\end{align}
which enables a relatively easy calculation of the first-passage time distribution in the case of single-reset systems. 

Let us now directly prove that in the single-reset systems the first-passage time distribution exhibits the renewal property (i.e. the subsequent first-passage times are uncorrelated). The general mathematical conditions of the renewal property can be defined as follows: The joint probability $F(N',N''|\tau',\tau'')$ that the jump number reaches the value $N'$ for in the moment $\tau'$ and than the number $N''$ in the moment $\tau''$ is a product of two first passage times distributions:
\begin{align}
F(N',N''|\tau',\tau'')=F(N'|\tau'')F(N''-N'|\tau''-\tau').
\end{align}
For the similar definition of the renewal property in the case of the waiting time distribution see Refs.~\cite{cao2006, gopich2006, budini2011}. As a result
\begin{align}
F(N''|\tau'')&=\int_0^{\tau''} F(N',N''|\tau',\tau'') d \tau' \\ \nonumber
& =\int_0^{\tau''} F(N'|\tau')F(N''-N'|\tau''-\tau') d \tau',
\end{align}
and therefore
\begin{align} \label{laplprod}
\hat{F}(N''|s) =\hat{F}(N'|s)\hat{F}(N''-N'|s),
\end{align}
when again the Laplace transform was applied. Equation~\eqref{laplprod} implies that $\hat{F}(N|s)=\hat{F}(N-1|s)\hat{F}(1|s)=\dots=\hat{F}(1|s)^N$. This is the necessary and sufficient condition of the renewal property. As one can observe, the relation $\hat{F}(N|s)=\hat{F}(1|s)^N$ directly results from Eq.~\eqref{fptdsingresodz} (with $\hat{F}(1|s)=[z_{+}(s)]^{-1}$). This proves that the first-passage time distribution in the singlet-reset systems exhibits the renewal property.

It is also apparent, that in renewal systems cumulants of the first-passage time distribution are linear functions of $N$:
\begin{align} \label{kumlinear}
\kappa^N_m=(-1)^m \lim_{s \rightarrow 0^{+}} \left[\frac{d^m}{ds^m} \ln \hat{F}(N|s) \right]=N \kappa^1_m,
\end{align}
where again the relation $\hat{F}(N|s)=\hat{F}(1|s)^N$ was used.

Furthermore, substituting $s \rightarrow g(z)$ in Eq.~\eqref{fptdsingresodz} and using the identity $z_{+}[g(z)]=z$ [see discussion below Eq.~\eqref{scfungen}] the following relations can be found:
\begin{align} \label{fodg1}
\hat{F}[N|g(z)]-z^{-N}&=0 \quad \text{for} \quad z>1, \\ \label{fodg2}
\ln \hat{F}[N|\lambda(l)]+N l&=0 \quad \text{for} \quad l>0,
\end{align}
where, as previously mentioned, $\lambda(l)=g(e^l)$. Since Eq.~\eqref{fodg2} implies that the expression $\ln \hat{F}[N|\lambda(l)]+N l$ is the constant function of $l$ for $l>0$, one obtains
\begin{align}
 \lim_{l \rightarrow 0^{+}}\left(\frac{d^m}{dl^m}\left\{ \ln \hat{F}[N|\lambda(l)]+N l \right\}  \right)=0.
\end{align}
Solving this equation for subsequent values of $m$, taking into account that due to $\lambda(0)=0$, ${\lambda'(0)>0}$~\cite{touchette2009} the following relation results from Eq.~\eqref{kumfpt}:
\begin{align}
\kappa^N_m=(-1)^m  \lim_{l \rightarrow 0^{+}} \left \{\frac{d^m}{d[\lambda(l)]^m} \ln \hat{F}[N|\lambda(l)] \right\},
\end{align}
and using definition of the scaled cumulants of the full counting statistics~\cite{bruderer2014}
\begin{align} \label{kumfcs}
c_m= \left[\frac{d^m}{dl^m} \lambda(l) \right]_{l =0},
\end{align}
one can obtain the relations between the cumulants of the full counting statistics and the first-passage time distribution. For example, the first three relations read as
\begin{align} \label{rel1}
c_1 &= \frac{N}{\kappa^N_1} = \frac{N}{\langle \tau_N \rangle},\\ \label{rel2}
\frac{c_2}{c_1}&= N \frac{\kappa^N_2}{(\kappa^N_1)^2} = N \frac{\langle \Delta \tau_N^2 \rangle}{\langle \tau_N \rangle^2}, \\ \label{rel3}
\frac{c_3}{c_1}&= N^2 \left[ 3 \frac{(\kappa^N_2)^2}{(\kappa^N_1)^4} - \frac{\kappa^N_3}{(\kappa^N_1)^3} \right],
\end{align}
where $\langle \tau_N \rangle$ and $\langle \Delta \tau_N^2 \rangle$ are the mean value and the variance of the distribution $F(N|\tau)$. This is the second main result of the paper. As mentioned, similar relations for cumulants of the waiting time distribution have been derived in Refs.~\cite{albert2011, budini2011}; here they are generalized to the case of bidirectional transitions. In Appexdix~\ref{sec:appendixb} I present an alternative, heuristic derivation of the relation between the first-passage time distribution and the full counting statistics based only on the renewal property. It may suggest that these relations apply also for general renewal systems.

It should be noted, that the relations between the first-passage time distribution and the full counting statistics have been already studied by Gingrich and Horowitz~\cite{gingrich2017}. However, the relations presented there were valid only in the limit of large threshold $N$; in contrast, the results presented here apply also to the case of low threshold values.

\section{\label{sec:nonrenewal}First-passage times in nonrenewal systems}
%%%%%%%%%%%%%%%%%%%%%%%%%%%%%%%%%%%%%%%%%%%%%%%%%%%%%%%%%%%%%%%%%%%%
\begin{figure}
	\centering
	\subfloat[]{\includegraphics[width=0.6\linewidth]{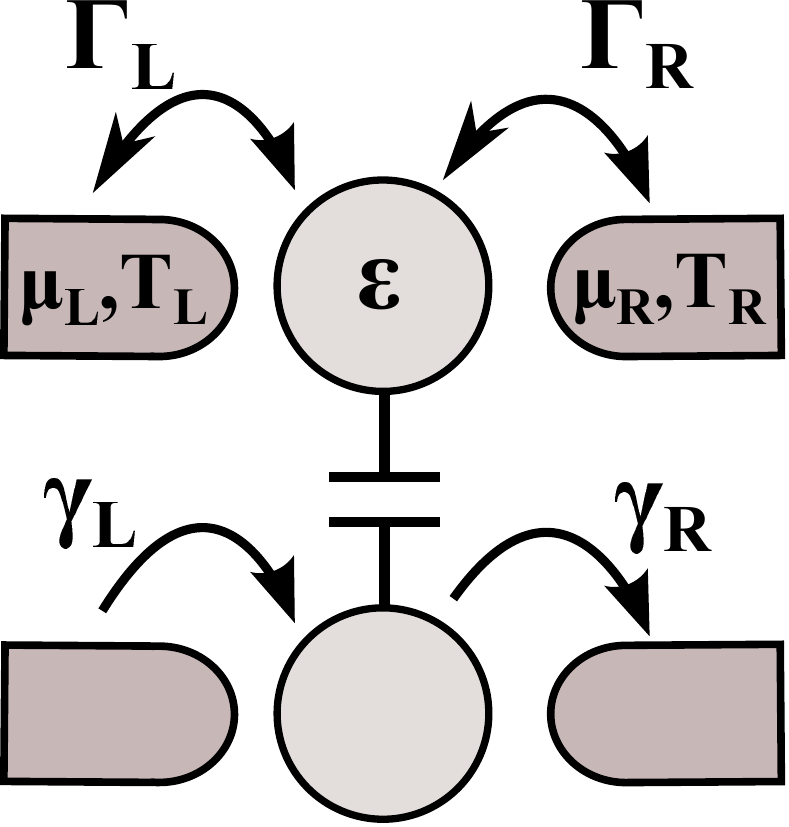}} \\
	\subfloat[]{\includegraphics[width=0.6\linewidth]{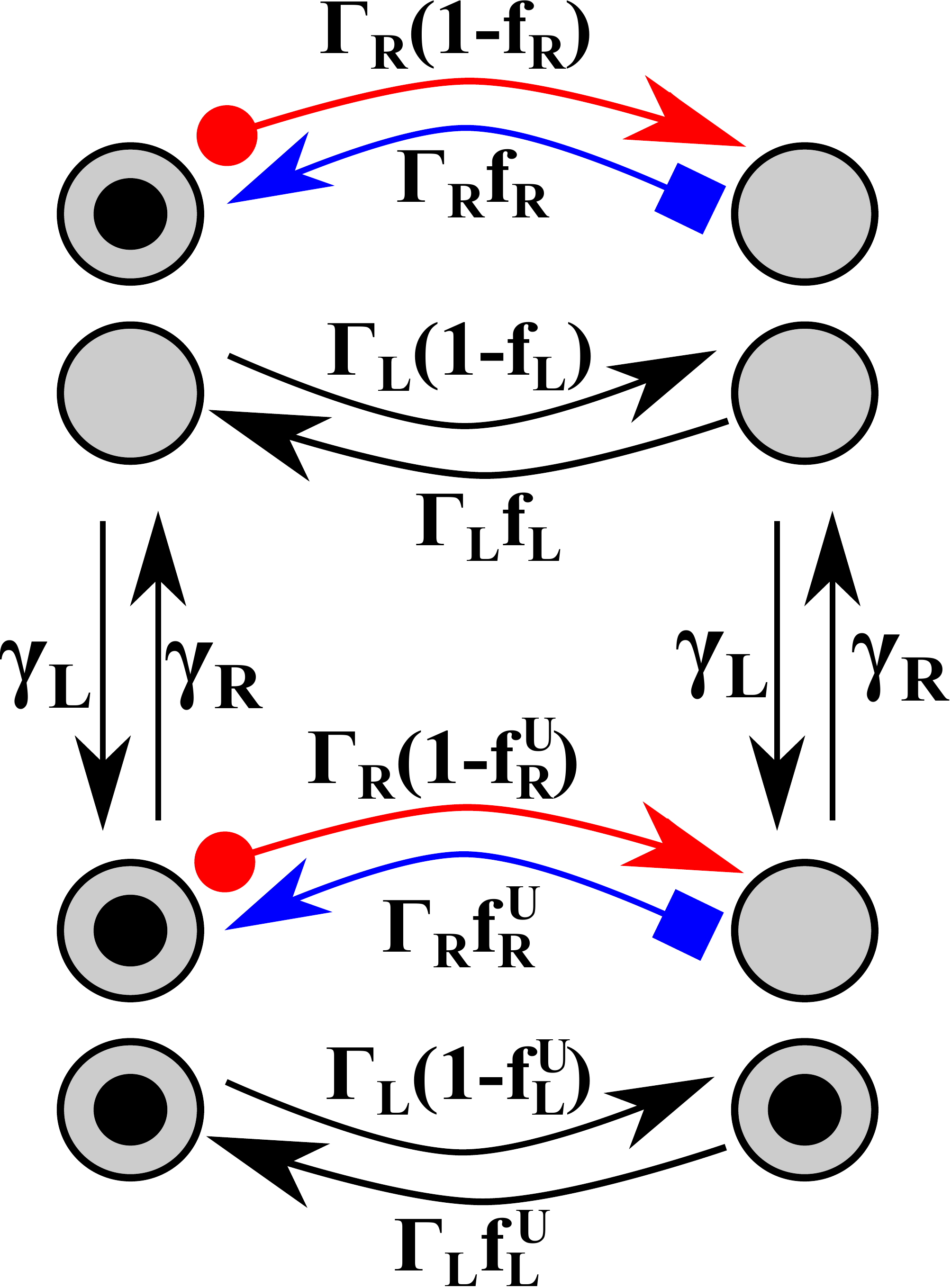}}
	\caption{(a) Scheme of the system of two capacitively coupled quantum dots, each attached to two leads. Orbital energy in the upper dot is equal to $\epsilon$. The electrochemical potentials and temperatures of the leads attached to the upper dot are equal to $\mu_\alpha$ and $T_\alpha$, respectively, with $\alpha \in \{L,R\}$. The transport in the lower dot is unidirectional, occuring from the left to the right lead, due to the high voltage bias. Inter-dot Coulomb interactions is equal to $U$. Tunneling rates in the upper and the lower dot, respectively, are denoted as $\Gamma_\alpha$ and $\gamma_\alpha$. (b) Four state model of the dynamics of the system. Terms $f_\alpha$ and $f_\alpha^U$ are Fermi distribution functions defined as $f_\alpha={f[(\epsilon-\mu_\alpha)/k_B T_\alpha]}$, $f_{\alpha}^U={f[(\epsilon+U-\mu_\alpha)/k_B T_\alpha]}$. Read arrows with bullet tails denote the forward transitions, and blue arrows with square tails denote the backward transitions.}
	\label{fig:dqdscheme}
\end{figure}
%%%%%%%%%%%%%%%%%%%%%%%%%%%%%%%%%%%%%%%%%%%%%%%%%%%%%%%%%%%%%%%%%%%%
%%%%%%%%%%%%%%%%%%%%%%%%%%%%%%%%%%%%%%%%%%%%%%%%%%%%%%%%%%%%%%%%%%%%
\begin{figure}
	\centering
	\includegraphics[width=0.9\linewidth]{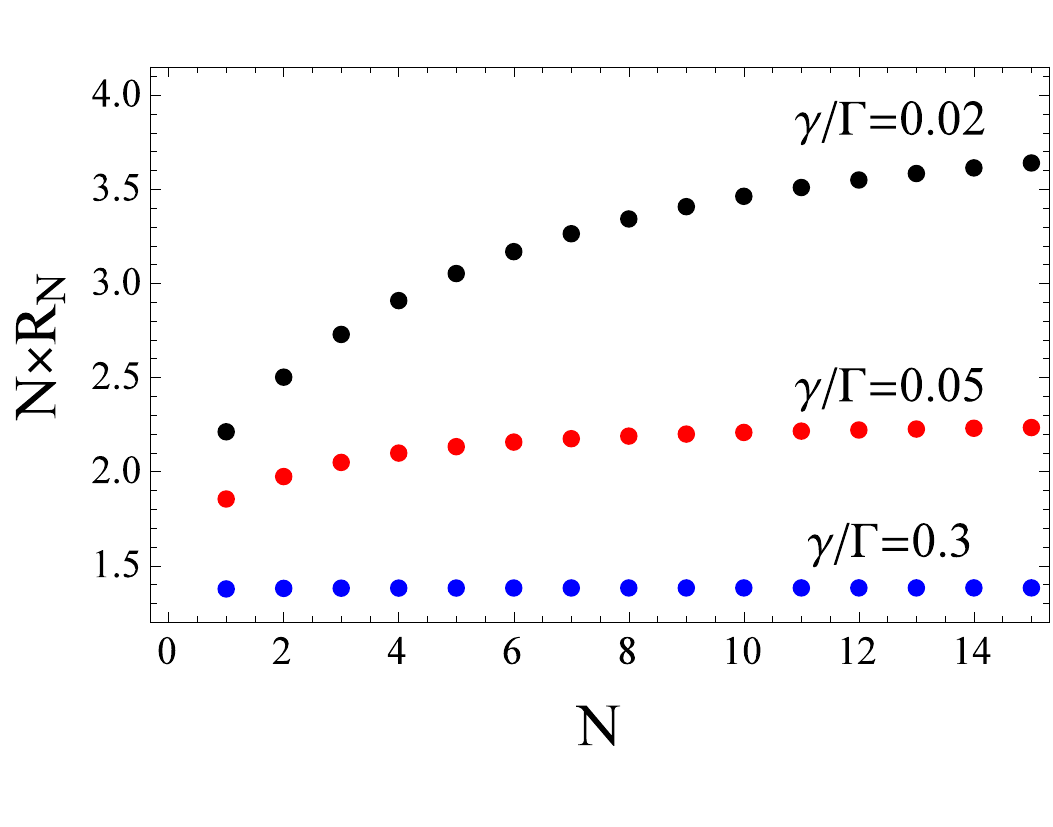}
	\caption{Randomness parameter $R_N$ multiplied by $N$ in the double dot system as a function of $N$ for different values of $\gamma/\Gamma$ (with $\gamma=\gamma_L=\gamma_R$ and $\Gamma=\Gamma_L=\Gamma_R$). All results for $\epsilon=0$, $\mu_L=1$, $\mu_R=-1$, $k_B T_L=k_B T_R=1$, $U=3$.}
	\label{fig:kumodn}
\end{figure}
%%%%%%%%%%%%%%%%%%%%%%%%%%%%%%%%%%%%%%%%%%%%%%%%%%%%%%%%%%%%%%%%%%%%
In this section, I consider a model multi-reset system in which the renewal property does not hold -- the subsequent first passage times are correlated. Presence of the correlations is directly confirmed by the analysis of cumulants of the distribution $F(N|\tau)$ for the subsequent values of $N$. Moreover, I show that in the multi-reset system the relation given by Eq.~\eqref{rel2} is violated. The joint analysis of the full counting statistics and the first passage time distribution may therefore reveal the multi-reset nature of the system.

To analyze the nonrenewal behavior, I consider a model system of two capacitively coupled quantum dots, each attached to two separate leads [Fig.~\ref{fig:dqdscheme}~(a)]. Current fluctuations in similar systems have been already studied both theoretically~\cite{michalek2009, schaller2010, cuetara2011} and experimentally~\cite{mcclure2007, singh2016}. In particular, a double dot system has been shown to exhibit the nonrenewal behavior in Ref.~\cite{ptaszynski2017}, where unidirectional electron transport has been analyzed. Here, bidirectional tunneling in the upper dot is enabled; for the sake of simplicity, the voltage bias in the lower dot is assumed to be large in comparison with $k_B T$, such that transport through this dot is unidirectional. I also assume, that the intradot Coulomb interaction is strong, such that only the zero- and the single occupancy of the dot is allowed. Furthermore, for the sake of simplicity, the spin is neglected; experimentally, the effectively spinless system can be obtained by applying a strong magnetic field, which breaks the degeneracy of the spin levels~\cite{keller2016}. Due to the Coulomb interaction between the dots, tunneling through the lower dot switches the charging energy in the upper dot between the two values: $\epsilon$ and $\epsilon+U$. This results in the switching between two values of the conductance of the upper dot. Such a phenomenon is often referred to as the telegraphic switching~\cite{ptaszynski2017}. 

In the weak tunnel coupling regime the system can be described by a Markovian master equation~\cite{michalek2009, schaller2010, cuetara2011}. The four state model of the dynamics of the system is shown in Fig.~\ref{fig:dqdscheme}~(b). One  may notice a similarity of the studied model to the ones describing the enzymatic networks with conformational fluctuations~\cite{barato2015, cao2011, kolomeisky2011}. The counted forward and backward transitions are associated with the tunneling between the upper dot and the upper left lead. As one can notice, each such transition is associated with two elementary transitions within the Markovian model, corresponding to different charge states of the lower dot.

I will focus on the analysis of two quantities, the randomness parameter $R_N$ and the Fano factor $FF$, which characterize the variances of the first-passage time distribution and the full counting statistics, respectively. They are defined as
\begin{align}
R_N &= \frac{\kappa^N_2}{(\kappa^N_1)^2}= \frac{\langle \Delta \tau_N^2 \rangle}{\langle \tau_N \rangle^2},  \\
FF &= \frac{c_2}{c_1}=\lim_{t \rightarrow \infty} \frac{\langle \Delta n(t)^2 \rangle}{\langle n(t) \rangle},
\end{align}
where $\langle n(t) \rangle$ and $\langle \Delta n(t)^2 \rangle$ are the mean value and the variance of the jump number $n(t)$ in the moment $t$. Due to the taken limit $t \rightarrow \infty$, the Fano factor characterizes the long time behavior of the current fluctuations. The randomness parameter, on the other hand, can characterize the short time dynamics of the system. 

Let us consider the dependence of the randomness parameter $R_N$ on $N$. Equation~\eqref{kumlinear} implies that in renewal systems $R_N=R_1/N$. As Fig.~\ref{fig:kumodn} shows, in nonrenewal systems this relation does not longer hold. This is associated with the presence of the correlation between subsequent waiting times. Variance of the distribution $F(2|\tau)$ is equal to
\begin{align}
\langle \Delta \tau_2^2 \rangle &= \langle (\tau_1+\tau'_1-2 \langle \tau_1 \rangle)^2 \rangle \\ \nonumber &= 2 \langle \Delta \tau_1^2 \rangle + 2 \langle \Delta \tau_1 \Delta \tau'_1 \rangle,
\end{align}
where $\tau_1$ and $\tau'_1$ are two subsequent first passage times for $N=1$. It is apparent, that it depends on the cross-correlation term $ 2 \langle \Delta \tau_1 \Delta \tau'_1 \rangle$. The randomness parameter $R_2$ can be then expressed as
\begin{align}
R_2=R_1 \frac{1+\text{NCC}}{2},
\end{align}
where
\begin{align}
\text{NCC}=\frac{\langle \Delta \tau_1 \Delta \tau'_1 \rangle}{\langle \Delta \tau_1^2 \rangle},
\end{align}
is the normalized cross-correlation of two subsequent first-passage times. For the analyzed system $R_2 \geq R_1/2$, and thus NCC is nonnegative. The positive cross-correlation of first-passage times results from the switching between the ``fast" and the ``slow" transport channels due to tunneling in the lower dot -- when the ``fast" channel is open two subsequent first-passage times tend to be shorter than the mean and vice versa. Similar behavior of waiting times in the case of unidirectional transport has been observed in Ref.~\cite{ptaszynski2017}.

%%%%%%%%%%%%%%%%%%%%%%%%%%%%%%%%%%%%%%%%%%%%%%%%%%%%%%%%%%%%%%%%%%%%
\begin{figure}
	\centering
	\subfloat{\includegraphics[width=0.9\linewidth]{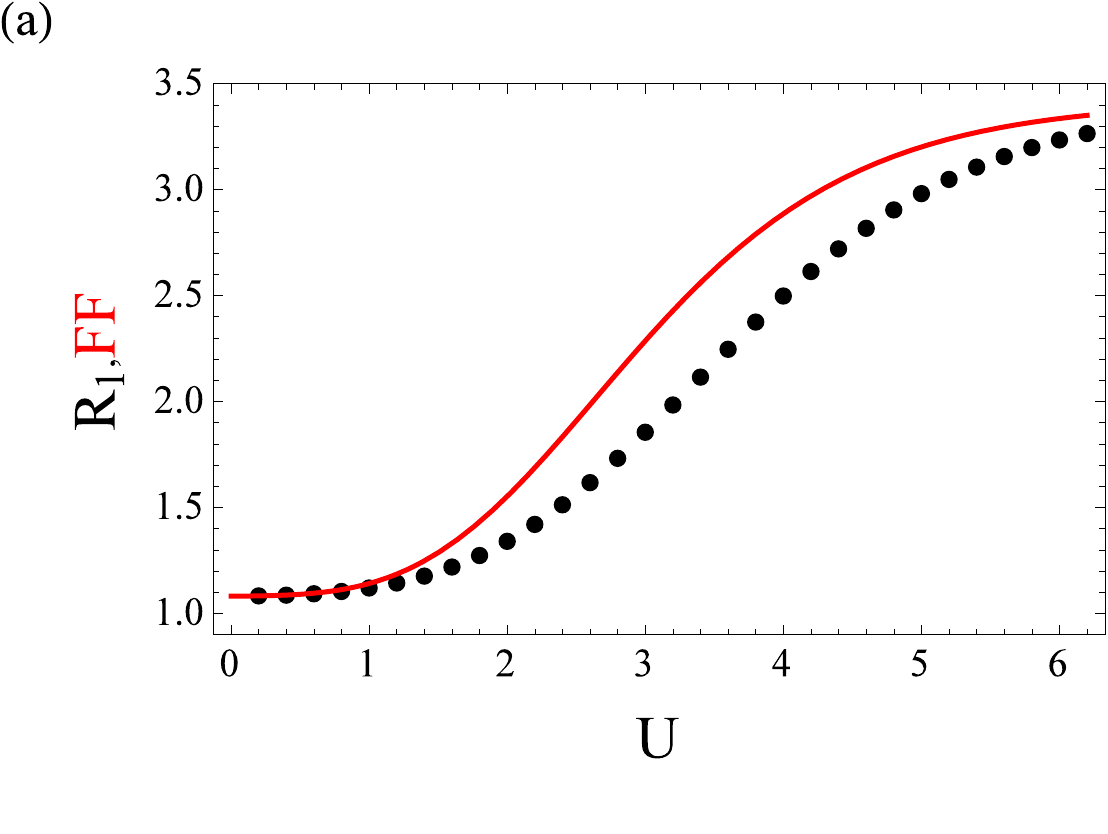}} \\
	\subfloat{\includegraphics[width=0.9\linewidth]{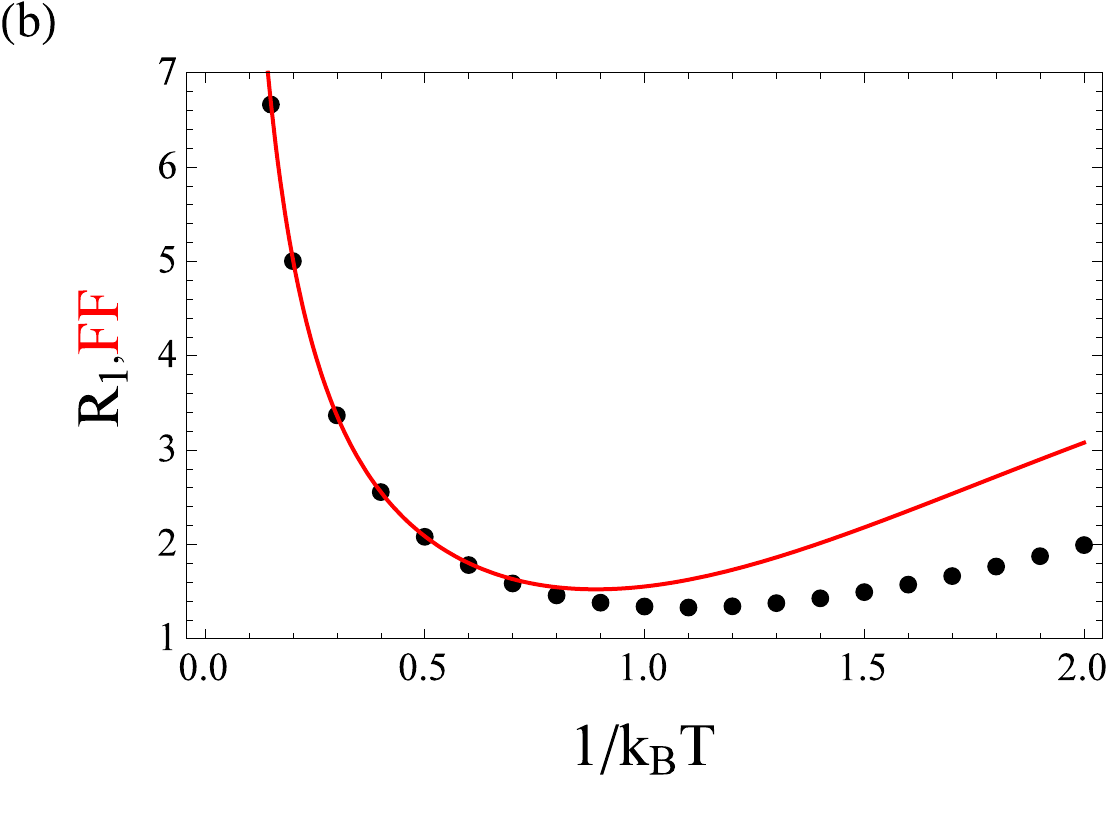}}
	\caption{Fano factor $FF$ (red solid line) and the randomness parameter $R_1$ (black dots) in the double dot system as a function of (a) $U$ for $k_B T_L=k_B T_R=1$, (b) $k_B T$ $(=k_B T_L=k_B T_R)$ for $U=2$. All results for $\epsilon=0$, $\mu_L=1$, $\mu_R=-1$, $\Gamma_L=\Gamma_R=1$, $\gamma_L=\gamma_R=0.05$.}
	\label{fig:fanorand}
\end{figure}
%%%%%%%%%%%%%%%%%%%%%%%%%%%%%%%%%%%%%%%%%%%%%%%%%%%%%%%%%%%%%%%%%%%%

As Eq.~\eqref{rel2} implies, in single-reset systems there exists a strict relation between the Fano factor and the randomness parameters. Figure~\ref{fig:fanorand} shows, that in the analyzed double-dot system this relation does not longer hold. Difference of the Fano factor $FF$ and the randomness parameter $R_1$ depends on the value of the intradot Coulomb interaction [Fig.~\ref{fig:fanorand}~(a)]. For $U=0$ both parameters are equal, since the transport in the upper dot is not affected by the dynamics of the lower dot. Difference is largest for intermediate values of $U$. For $U \rightarrow \infty$, parameters become equal again, because transport is completely blocked when the lower dot is occupied, and the system is again renewal -- there is no transport through a ``slow'' channel, and therefore there are no series of subsequent ``long'' first-passage times.

Dependence of the analyzed quantities on the temperature is shown in Fig.~\ref{fig:fanorand}~(b). One can observe that for $k_B T>U$ the Fano factor $FF$ and the randomness parameter $R_1$ are approximately equal. This results from strong thermal fluctuations which mask the influence of the Coulomb interaction on the transport. Current fluctuations in this regime result mainly from the thermal noise and the telegraphic switching is not observed. Inequality of the Fano factor and the randomness parameter can be observed for $k_B T<U$, when current fluctuations are strongly affected by the non-thermal effects like the telegraphic switching.

\section{\label{sec:fluctuation}Violation of the fluctuation theorem in multicyclic systems}
%%%%%%%%%%%%%%%%%%%%%%%%%%%%%%%%%%%%%%%%%%%%%%%%%%%%%%%%%%%%%%%%%%%%
\begin{figure}
	\centering
	\subfloat[]{\includegraphics[width=0.6\linewidth]{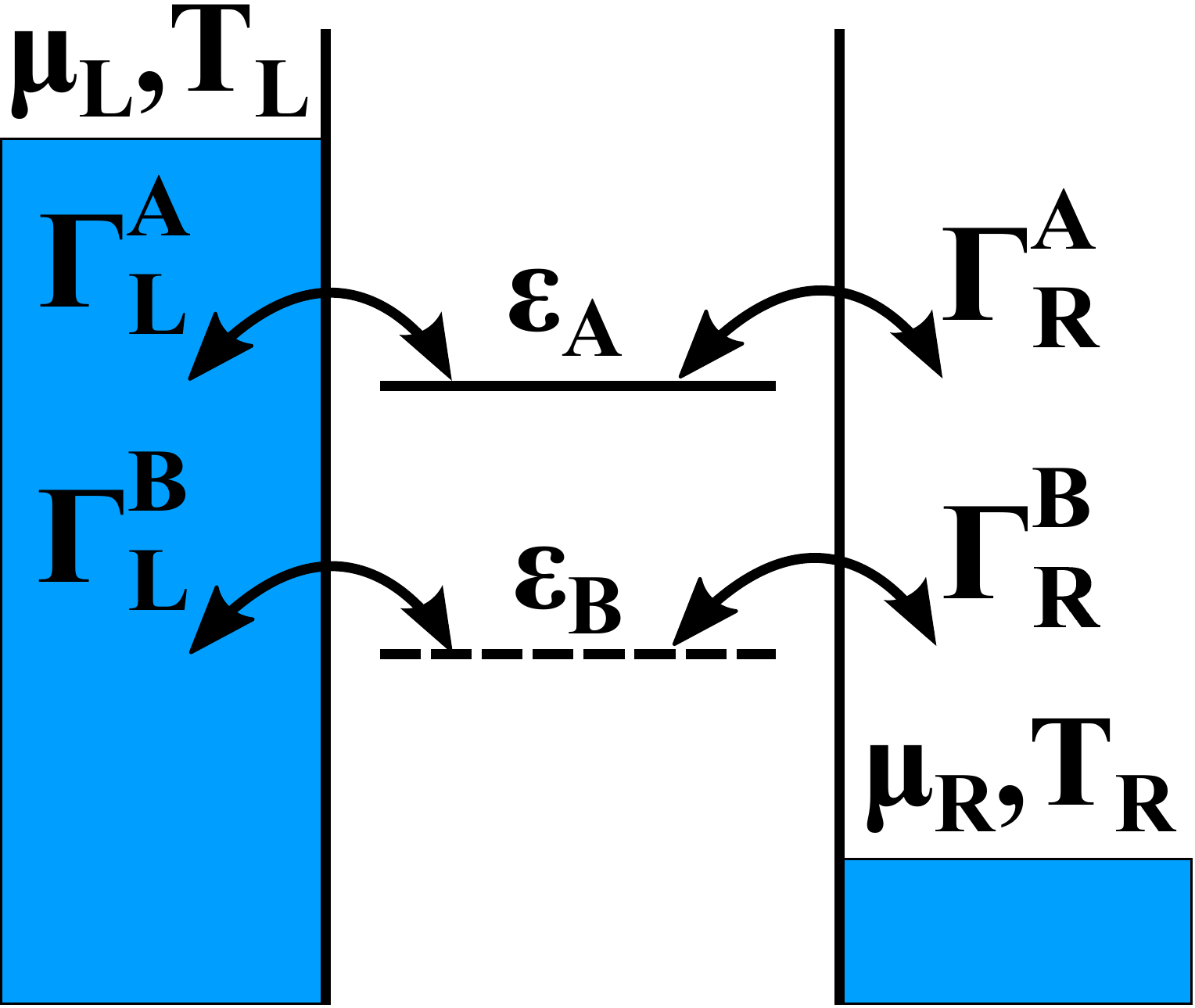}} \\
	\subfloat[]{\includegraphics[width=0.9\linewidth]{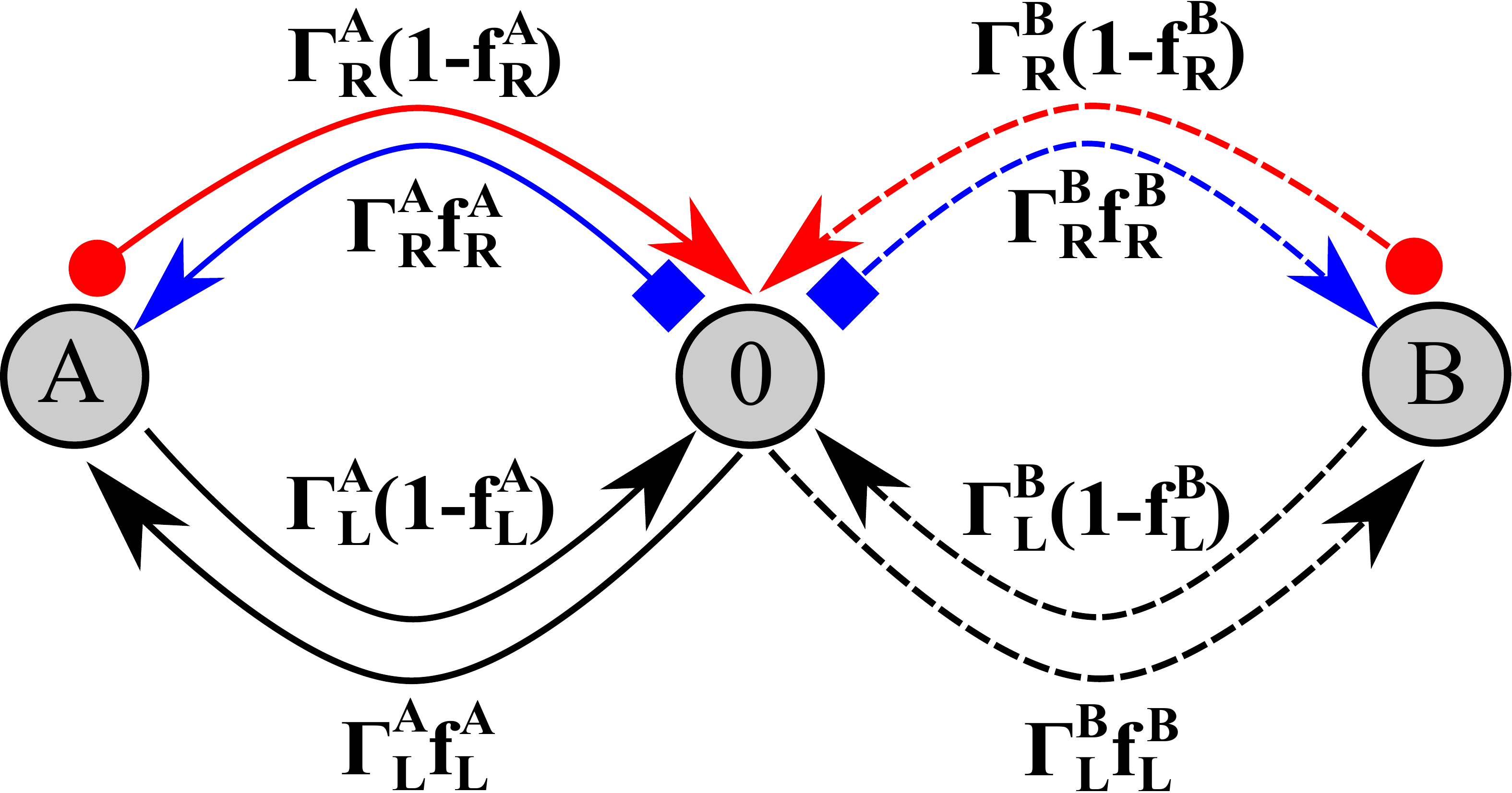}}
	\caption{(a) Scheme of the transport through a single quantum dot. The cases when a single level A or two levels A and B are available for the transport are considered. Energies of the dot levels $A$ and $B$ are equal to $\epsilon_A$ and $\epsilon_B$. The dot is attached to two leads with electrochemical potentials $\mu_\alpha$ and temperatures $T_\alpha$ with $\alpha \in \{L,R\}$. Tunneling rates between the leads and specific levels are denoted as $\Gamma^i_\alpha$ with $i \in \{A,B\}$. (b) Three state Markovian model of the dynamics of the system. Terms $f_\alpha^i$ are the Fermi distribution functions defined as $f_{\alpha}^i=f[(\epsilon_i-\mu_\alpha)/k_B T_\alpha]$. Solid/dashed lines correspond to the tunneling through the level $A$/$B$. Red arrows with bullet tails denote the forward transitions while blue arrows with square tails denote the backward transitions.}
	\label{fig:dcbscheme}
\end{figure}
%%%%%%%%%%%%%%%%%%%%%%%%%%%%%%%%%%%%%%%%%%%%%%%%%%%%%%%%%%%%%%%%%%%%
%%%%%%%%%%%%%%%%%%%%%%%%%%%%%%%%%%%%%%%%%%%%%%%%%%%%%%%%%%%%%%%%%%%%
\begin{figure}
	\centering
	\subfloat{\includegraphics[width=0.9\linewidth]{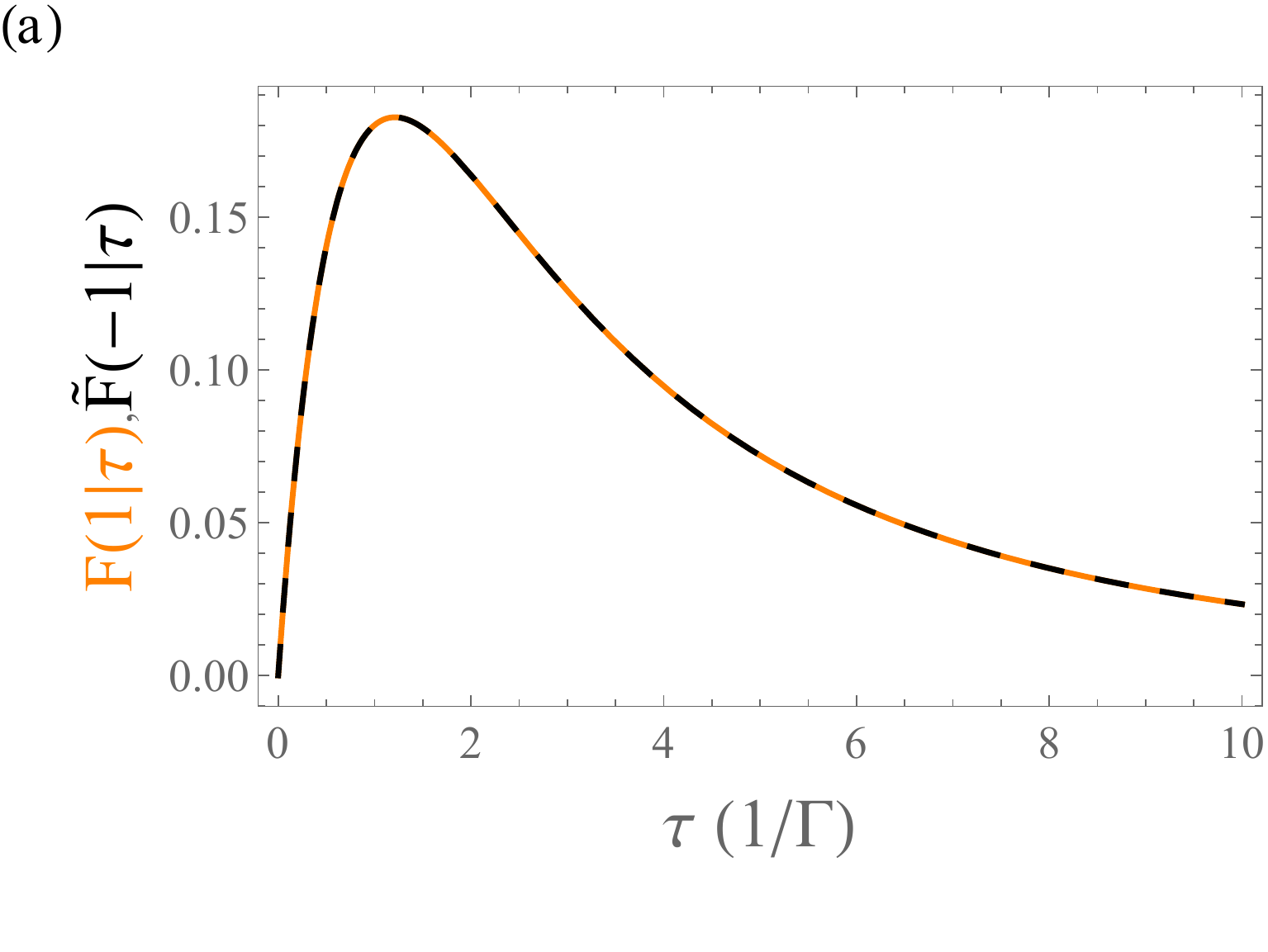}} \\
	\subfloat{\includegraphics[width=0.9\linewidth]{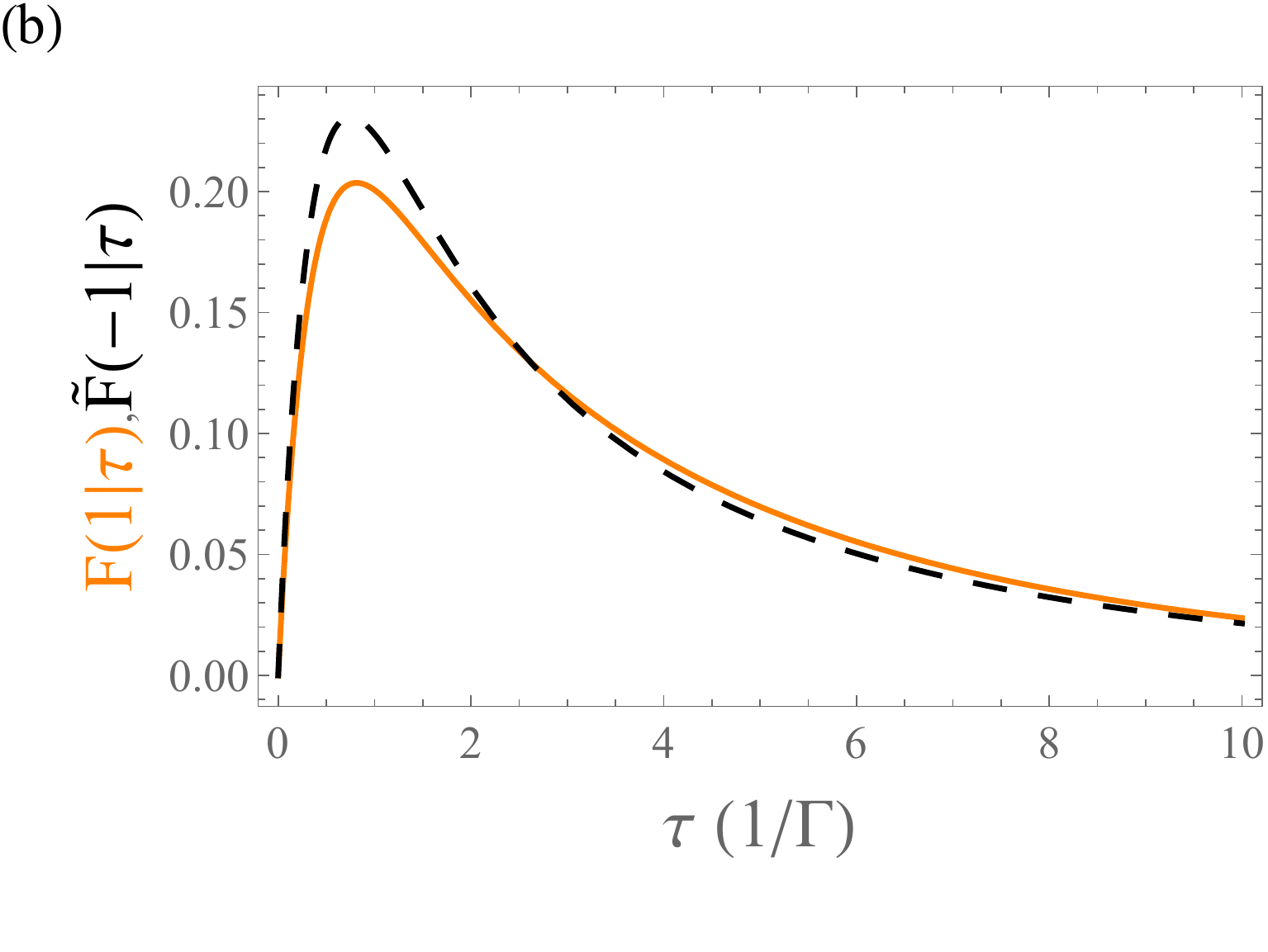}}
	\caption{First-passage time distribution $F(1|\tau)$ (orange solid line) and the normalized first-passage time distribution $\tilde{F}(-1|\tau)=F(-1|\tau)/[\int_0^\infty F(-1|\tau) d\tau]$ (black dashed line) for (a) a single level dot with $\epsilon_A=0$ and $\Gamma_L^A=\Gamma_R^A=\Gamma$, (b) a two-level dot with $\epsilon_A=0$, $\epsilon_B=-2$, $\Gamma_L^A=\Gamma_R^A=\Gamma_L^B=\Gamma_L^B=\Gamma$. All results for $\mu_L=1$, $\mu_R=-1$, $k_B T_L=1$, $k_B T_R=2$.}
	\label{fig:flucttheor}
\end{figure}
%%%%%%%%%%%%%%%%%%%%%%%%%%%%%%%%%%%%%%%%%%%%%%%%%%%%%%%%%%%%%%%%%%%%
As mentioned in Sec.~\ref{sec:intro}, the current fluctuation statistics can be used to infer the structure and dynamics of the Markovian network. The previous sections have shown that the first-passage time distribution can be used to distinguish between renewal and non-renewal systems; this may enable, for example, to infer the presence of the switching mechanism. This sections shows that it can be also applied to distinguish between unicyclic and multicyclic systems. These terms have the following meaning: According to the decomposition scheme of Schnakenberg~\cite{schnakenberg1976}, in each Markovian network one can identify a complete set of fundamental cycles. Each cycle is associated with an affinity $\mathcal{A}_\beta$. If there is only one fundamental cycle, the system is referred to as a unicyclic one; otherwise, it is a multicyclic one. In the case of unidirectional transitions it has been shown that multicyclic nature of the system can be inferred by analysis of higher moments of the waiting time distribution~\cite{barato2015prlb}. Here I show that in the case of bidirectional transition this can be revealed by breaking of the fluctuation theorem valid for unicyclic systems, derived by Bauer and Cornu~\cite{bauer2014}.

The difference between the unicyclic and the multicyclic systems will be discussed on the basis of the model of transport through a single spinless quantum-dot in a strong Coulomb blockade regime (i.e. with only a single occupancy of the dot allowed), with either one or two levels [Fig.~\ref{fig:dcbscheme}~(a)]. As a Markovian model of the dynamics presented in Fig.~\ref{fig:dcbscheme}~(b) shows, tunneling through a single-level dot (with only the level A available for the transport) is associated with a single thermodynamic cycle (denoted by the solid lines). Upper (lower) branch of the cycle is associated with tunneling through the right (left) lead. On the other hand, tunneling through a two-level quantum dot is described by a bicyclic model, with separate cycles corresponding to tunneling through the levels A (solid lines) and B (dashed lines). On can notice that both systems are single-reset ones, since after each tunneling from the dot the system returns to the same state 0.

First, I focus on the case of the unicyclic single-level dot system. Here the forward and the backward transitions correspond to the jumps in the clockwise and the anticlockwise direction within the cycle. The first-passage time distributions for forward and backward transitions are then related by the fluctuation theorem of Bauer and Cornu~\cite{bauer2014} 
\begin{align} \label{flucttheor}
\frac{F(1|\tau)}{F(-1|\tau)}=e^{\mathcal{A}},
\end{align}
where $\mathcal{A}$ is the affinity of the cycle measured in the clockwise direction [here $\mathcal{A}={(\mu_L-\epsilon_A)/k_B T_L}+{(\epsilon_A-\mu_R)/k_B T_R}$]. Figure~\ref{fig:flucttheor}~(a) illustrates the validity of this theorem.

As Fig.~\ref{fig:flucttheor}~(b) shows, in the case of the two-level dot distributions $F(1|\tau)$ and $F(-1|\tau)$ are not proportional to each other any longer [$F(1|\tau)/F(-1|\tau) \neq \mathrm{const}$]. This results from the fact that the counted transition corresponds now to jumps in two distinct cycles associated with different affinities. Violation of the fluctuation theorem for the  first-passage times given by Eq.~\eqref{flucttheor} may therefore reveal the multicyclic character of the system. It should be noted, that this can be also inferred from breaking of the Gallavotti-Cohen symmetry for non-entropic currents~\cite{barato2012}.

\section{\label{sec:conclusions}Conclusions}

The first-passage time distribution, i.e. the distribution of time delays after which the measure quantity reaches some target value, has been studied in systems described by discrete Markovian networks by means of the master equation. In Sec.~\ref{sec:method} the equation enabling to determine the first-passage time distribution for currents associated with arbitrary sets of transitions within the system has been derived. In Secs.~\ref{sec:singleres}-\ref{sec:nonrenewal} this equation has been applied to study the relation between the first-passage time distribution and the full counting statistics in system in which the subsequent first-passage times are either correlated or uncorrelated (referred to as renewal and nonrenewal systems, respectively). In single-reset systems, which are renewal ones, the cumulants of the first-passage time distribution are shown to be related to the cumulants of the full counting statistics. The obtained relations do not longer hold in the case of nonrenewal systems. Moreover, correlations between subsequent first-passage times can be investigated by measuring cumulants of the first-passage time distribution for different target values.

Furthermore, in Sec.~\ref{sec:fluctuation} behavior of the first-passage time distribution in unicyclic and multicyclic systems has been investigated. In unicyclic systems the fluctuation theorem holds, which relates the first-passage time distributions for target values of the different sign. In multicyclic systems this theorem, in general, does not longer hold. Therefore, the first-passage time distribution may be used to infer the multicyclic nature of the Markovian network.

The first-passage time distribution may be therefore a useful tool to characterize the statistical kinetics of biomolecular reactions or electronic transport in mesoscopic systems. It seems to be particularly valuable in the case of nonrenewal systems, when it provides additional information in comparison to the one provided by the full counting statistics. Analysis of the nonrenewal behavior may reveal and characterize the hidden internal dynamics of the system, which can be associated, for example, with the switching between the conformational states of the molecule~\cite{gopich2006, cao2011, kolomeisky2011} or charge~\cite{fricke2007, ptaszynski2017}, spin~\cite{sothmann2010, ptaszynski2017b} or phonon~\cite{koch2005, lau2016, kosov2017} dynamics in electronic systems.

There are still some open issues. For example, cumulants of the full counting statistics have been shown to provide bounds on a minimal entropy production in the system~\cite{barato2015prl, barato2015, gingrich2016, pietzonka2016, polettini2016, pietzonka2017, horowitz2017}. Similar bound was also derived for the first-passage time distribution in the limit of large threshold; it is equivalent to the one provided by the zero-frequency full counting statistics~\cite{gingrich2017}. It would be worthwhile to check, if in nonrenewal systems similar but independent (and possibly tighter) bounds can be provided by the cumulants of the first-passage time distribution for an arbitrary threshold. However, due to the technical difficulty of determining the analytical form of the first-passage time distribution in multi-reset systems (associated with the necessity of calculating the complex integrals), derivation of such bounds represents a serious mathematical challenge.

\section*{Acknowledgments}
I thank B. R. Bu\l{}ka for the careful reading of the manuscript and the valuable discussion. This work has been supported by the National Science Centre, Poland, under the project 2016/21/B/ST3/02160. 

\appendix

\section{\label{sec:appendix}Derivation of Eq.~(\ref{detpolyn})}
First, let us chose $\nu=M$, where $M$ is the rank of the matrix $\mathbf{W}_z$. The matrix ${s-\mathbf{W}_z}$ can be then written in the block form
\begin{align}
s-\mathbf{W}_z=
\begin{pmatrix}
\mathbf{U} & \mathbf{Z}_{-} \\
\mathbf{Z}_{+} &  s-W_{MM} 
\end{pmatrix},
\end{align}
where $W_{MM}=-\sum_{i<M} k_{iM}$ and $\mathbf{U}$ is the matrix of rank $M-1$. Only $\mathbf{Z}_+$ and $\mathbf{Z}_-$ matrix vectors contain elements proportional to $z$ or $z^{-1}$. Using the properties of the determinants of the block matrices~\cite{dym2013} one obtains
\begin{align}
&\det[s-\mathbf{W}_z] =\det \mathbf{U} \det[s-W_{MM}-\mathbf{Z}_{+} \mathbf{U}^{-1} \mathbf{Z}_{-}] \nonumber
\\ \nonumber &= \det \mathbf{U} \left\{ s-W_{MM} \right. 
\\ & -\sum_{i,j<M} \left. \left(k_{Mi}^F z+k_{Mi}^R \right)\left(k_{jM}^B/ z+k_{jM}^R \right) V_{ij}\right\},
\end{align}
where terms $V_{ij}$ are elements of the matrix $\mathbf{V}=\mathbf{U}^{-1}$, $k_{Mi}^F +k_{Mi}^R=k_{Mi}$ and $k_{jM}^B+k_{jM}^R=k_{jM}$. This expression contains only elements proportional to $z$, $z^0$ and $z^{-1}$, and therefore can be expressed as in Eq.~\eqref{detpolyn}.

\section{\label{sec:appendixb} Alternative derivation of Eq.~(\ref{fodg1})}
Here I provide a heuristic argumentation for the applicability of Eqs.~\eqref{rel1}-\eqref{rel3} to any systems which exhibit the renewal property. First, I notice that if the transitions in the forward direction are more probable than the reverse process, for sufficiently large times $n$-conditioned probabilities $p^{(n)}(t)=\text{Tr}[\mathbf{p}^{(n)}(t)]$ for $n<0$ can be neglected, and the generating function $p(z,t)=\text{Tr}[\mathbf{p}(z,t)]$ can be expressed as
\begin{align} \label{apbfungen}
p(z,t) = \sum_{n=0}^\infty z^n p^{(n)}(t).
\end{align}
Secondly, I assume that probabilities $p^{(n)}(t)$ for $n>0$ can be expressed by equation similar to Eq.~\eqref{integraleq1lapl}:
\begin{align}
p^{(n)}(t)=\int_0^t F(n|\tau) p^{(0)}(t-\tau) d \tau.
\end{align} 
This assumption may be considered as a definition of the renewal property. Then I apply Laplace transform to Eq.~\eqref{apbfungen} and use the property $\hat{F}(n|s)=\hat{F}(1|s)^n$. In this way one obtains
\begin{align} \label{apbcond1}
p(z,s) &= \sum_{n=0}^\infty z^n p^{(n)}(s)= \sum_{n=0}^\infty p^{(0)}(s) \hat{F}(1|s)^n z^n \\ \nonumber &= \frac{p^{(0)}(s)}{1-\hat{F}(1|s)z}.
\end{align}
On the other hand, for sufficiently large times $p(z,t) = \exp [g(z)t]$~\cite{touchette2009}. Applying the Laplace transform one obtains
\begin{align} \label{apbcond2}
p(z,s)=\frac{1}{s-g(z)}.
\end{align}
Now I look for a condition in which both expressions for $p(z,s)$ given by Eqs.~\eqref{apbcond1} and~\eqref{apbcond2} are equivalent. I make a bold assumption: the condition is met when both expressions are singular, i.e. when $s=g(z)$ and ${1-\hat{F}(1|s)z}={1-\hat{F}[1|g(z)]z}=0$. This lead to Eq.~\eqref{fodg1} from which Eqs.~\eqref{rel1}-\eqref{rel3} can be derived. 

It should be noticed that during the derivation I made the assumptions which are not easy to justify. For example the Laplace transform is applied to functions which approximate $p(z,t)$ for large times; the Laplace transform, on the other hand, involves the integration over the whole time domain. Therefore, this heuristic argument should not be considered as a formal proof.

\end{document}